\documentclass[12pt]{article}
\usepackage{epsfig}
\begin{document}
\begin{titlepage}
\begin{flushright}
Cavendish-HEP-98/17\\
hep-ph/9811478\\
\end{flushright}
\vspace*{\fill}
\begin{center}
{\Large \bf Effects of Perturbative Colour Interference on 
$WW\rightarrow \mbox{6-Jet}$ Distributions}
\end{center}
\par \vskip 5mm
\begin{center}
	{\large Mark Smith}\footnote{Research supported
by U.K Particle Physics and Astronomy Research Council}\\
	Cavendish Laboratory, University of Cambridge,\\
	Madingley Road, Cambridge CB3 0HE, UK\\
	26 November 1998\\
\end{center}
\par \vskip 2mm
\begin{center} {\large \bf Abstract} \end{center}
\begin{quote}
At LEP II it is hoped to measure the W mass to an accuracy of around 40 MeV. 
This will require direct reconstruction of the mass of the W from its decay 
products in both the semi-leptonic and hadronic decay channels.  
Final state perturbative reconnection effects in hadronic decays are 
considered and their effect on 6-jet distributions and the reconstructed mass.
The perturbative mass shift is found to be $\sim$ 50 keV in the negative 
direction.
\end{quote}
\vspace*{\fill}
\end{titlepage}

\newpage

\section{Introduction}

One of the main goals of LEP II will be an accurate determination of the 
mass of the W boson.  An integrated luminosity of $500 \mbox{ pb}^{-1}$
suggests that an accuracy of $30 - 50 \mbox{ MeV}$ \cite{1} could be 
reached.  The process
\begin{equation}
e^{+}e^{-} \rightarrow W^{+}W^{-} \rightarrow \quad
\mbox{4 fermions}
\end{equation}
can be split into three distinct classes 
depending on the type of decay of each
W-boson.

\begin{itemize}
	\item Purely leptonic.  Both W-bosons decay to leptons.
	There are two neutrinos and reconstruction of the event from
	observed charged lepton momenta is not possible.  Branching ratio
	for this channel $\sim \frac{1}{9}$.

	\item Semi-leptonic.  One W-boson decays to leptons, the other
	decays hadronically.  One neutrino is produced, but the missing 
	momentum can be reconstructed using energy-momentum conservation
	and assumptions about the initial state radiation.  Branching ratio
	for this channel $\sim \frac{4}{9}$.

	\item Fully hadronic.  Both W-bosons decay hadronically.  All
	momenta are observable.  The momenta directions are well resolved,
	while the energy resolution can be improved via kinematic fits (ie
	imposing the constraints of energy and momentum conservation).
	Branching ratio for this channel $\sim \frac{4}{9}$.

\end{itemize}

In order to achieve the greatest accuracy the W mass must be reconstructed 
using both the semi-leptonic and the fully hadronic decay channels. 
However W's decay very rapidly so one expects that the 
space-time separation of the two decays should be $\sim 0.1 \mbox{ fm}$.  This
is small compared to the typical scale of 
hadronization $\sim 1 \mbox{ fm}$, thus
in the case of fully hadronic decay there are two evolving hadronic systems 
with considerable space-time overlap.  There is the possibility that
the two systems do not evolve independently, but influence each other.  

These influences fall into two categories\footnote{I neglect the effects
of electroweak interactions between the two systems as these have been
considered elsewhere\cite{2,3}}
 - Bose-Einstein correlations
between identical bosons in the final state (typically pions)
\cite{4,5,6}, and a re-arrangement of the colour flow of the evolving systems
at either the perturbative or 
hadronization level\cite{7,8,9,10}.  There has been much 
work on the effects of colour re-arrangement at the 
hadronization level, however hadronization is poorly understood and 
progress can only be made through constructing models.  It is interesting to 
note that the models of colour reconnections in the hadronization phase
give rather varied predictions\cite{1,11} for the 
effects on physical observables such as
mean charged multiplicity or reconstructed W mass, and so such measurements
may probe directly aspects of the confinement mechanism.

\begin{figure}
\centering
\epsfig{file=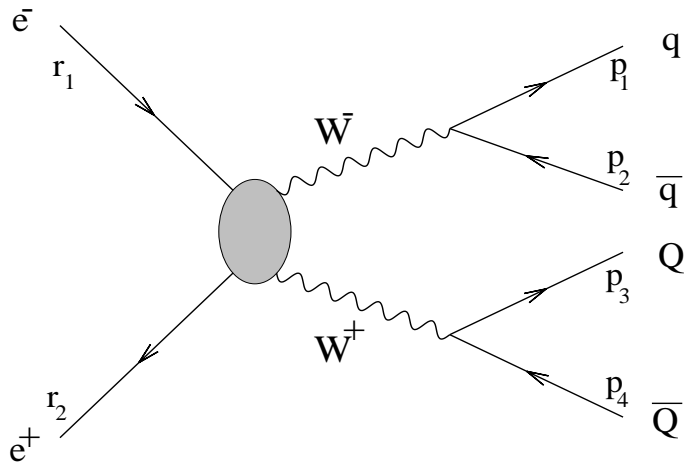,height=6cm,width=9cm}
\caption{$e^{+}e^{-}\rightarrow W^{+}W^{-}\rightarrow q \bar{q} Q \bar{Q}$.
The blob represents a sum over the three lowest order production amplitudes.}
\end{figure}

%\begin{figure}
%\epsfig{file=fig6.eps,height=4cm,width=13.5cm}
%\caption{Lowest order amplitudes for $e^{+}e^{-}\rightarrow
%W^{+}W^{-}\rightarrow q\bar{q} q\bar{q}$}
%\end{figure}

In this paper I will examine the 
effects of colour reconnection at its lowest non-trivial
order in perturbation theory.  In section two I will explain why 
these effects should be small and how they can be  
calculated directly.  In section three I shall present results for the effects
of colour reconnection on various distributions including the W mass.  The 
conclusions will be found in section four.

\section{Perturbative Reconnection}

\begin{figure}
\centering
\epsfig{file=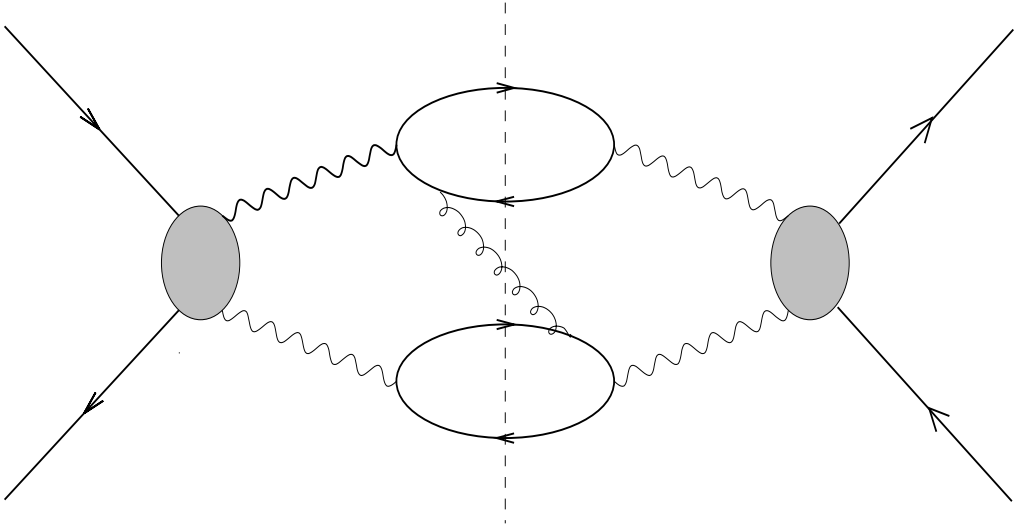,height=7cm,width=13.5cm}
\caption{A possible interference term involving one exchanged gluon (the 
shaded blobs represent a sum over the three W-pair production 
amplitudes).}
\end{figure}

Perturbative reconnection appears as higher order corrections to the process
shown in fig 1, 
in which gluons are exchanged between the evolving quark systems.  A 
possible reconnection diagram is shown in fig 2; here a real gluon emitted
from one decay system interferes with the similar emission from the other 
decay system.  One may also consider the analogous virtual interference 
corresponding to the exchange of a virtual gluon between decay systems.  
Within perturbation theory these interference terms are zero due to colour 
conservation\footnote{However it is not impossible for a colour octet
to be exchanged between the two decay systems at the perturbative level, 
only to be balanced by a non-perturbative exchange in the hadronization phase.
Such interplay between perturbative and non-perturbative connections is not 
considered here.}, and so one must consider the exchange of at least two 
perturbative gluons.

A full calculation of the $\mathcal{O}(\alpha_{s}^{2})$ 
corrections is beyond
the scope of this paper, however it is possible to examine QCD interference
effects in the production of 6 jets\cite{12} via
\begin{equation}
	e^{+}e^{-}\rightarrow W^{+}W^{-}\rightarrow q\bar{q}q\bar{q}gg
\end{equation}
in which interferences appear between the lowest order diagrams.  Two possible 
interference terms are shown in fig 3.

\begin{figure}
\centering
\epsfig{file=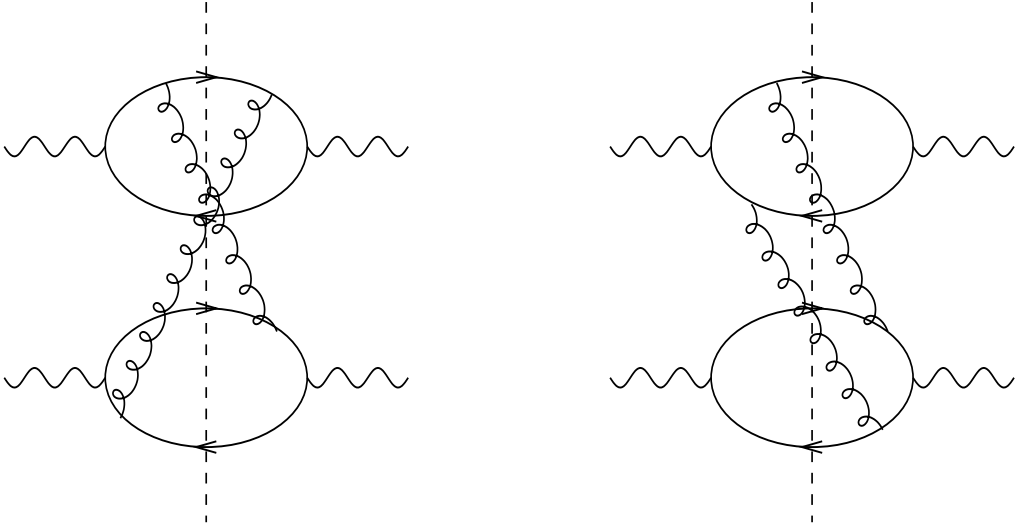,height=7cm,width=13.5cm}
\caption{Interference terms involving the exchange of two gluons (the W-pair
production parts of the diagrams are omitted for clarity)}
\end{figure}

These diagrams contain only two colour loops, compared with the diagrams for
gluon emissions within each decay system which contain four loops.  Therefore
the interference terms are suppressed relative to the leading emission by
$\frac{1}{N_{C}^{2}}$ where $N_{C}$ is the number of colours.

There is further suppression due to the width of the W.  Gluons radiated within
a decay system are free to have any energy up to $\sim M_{W}$ without
pushing the W Breit-Wigner propagators off resonance.  However gluons radiated
between the decay systems (interference terms) must carry energy less than 
$\sim \Gamma_{W}$ or at least two of 
the W propagators must be pushed off
resonance and that term will become suppressed.  It has been shown that for
inclusive quantities, where the I.R divergences cancel between real and 
virtual diagrams, that this leads to a suppression of perturbative 
reconnection effects by 
$\mathcal{O}(\frac{\Gamma_{W}}{M_{W}})$\cite{13,15}.
A rough estimate of the size of perturbative reconnection effects in W-pair
production is thus:

\begin{equation}
\frac{\Delta \sigma}{\sigma_{0}} \sim \frac{\alpha_{s}^{2}}{N_{C}^{2}}
\frac{\Gamma_{W}}{M_{W}} \sim 10^{-4}
\end{equation}
and so one may estimate the possible mass shift as
\begin{equation}
\Delta M_{W} \sim \frac{\alpha_{s}^{2}}{N_{C}^{2}} \Gamma_{W}
\sim \mbox{a few MeV}
\end {equation}

This should be regarded as an order of magnitude estimate only.  It is clearly
desirable to calculate experimental distributions in fixed order perturbation
theory and examine how they may be distorted by the effects of colour 
interference.

Using the helicity methods of \cite{15} it is possible to construct the
amplitudes for all the doubly resonant diagrams contributing to the 
$q\bar{q}q\bar{q}gg$ final state\footnote{Strictly speaking this is not a 
gauge invariant set of diagrams, however one may see that a change of gauge 
leads to singly resonant contributions which are neglected.  
The amplitudes were evaluated in the unitary gauge.}; there are 72
diagrams in total.  In this method each amplitude is built up from 
relatively few component pieces
that can be calculated separately.  
For example the amplitude for fig 1 can be written

\begin{figure}
\epsfig{file=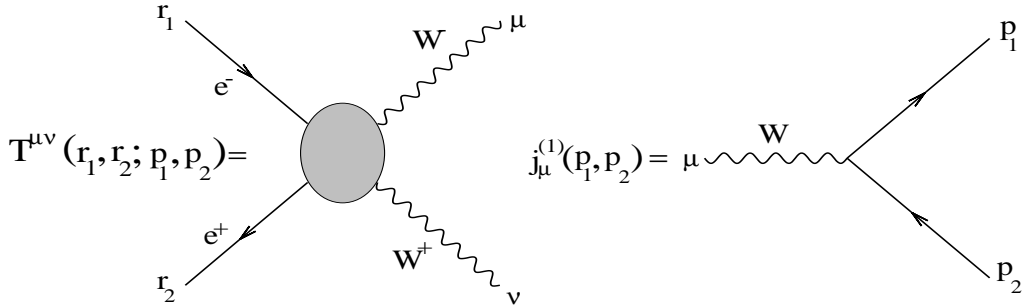,height=4cm,width=13.5cm}
\caption{Components used to calculate lowest order $e^{+}e^{-}\rightarrow
W^{+}W^{-}\rightarrow q\bar{q} q\bar{q}$.  The shaded blob represents the 
sum of the three interfering amplitudes, and the WW propagators have been 
absorbed into the definition of $T^{\mu \nu}$}
\end{figure}

\begin{figure}
\epsfig{file=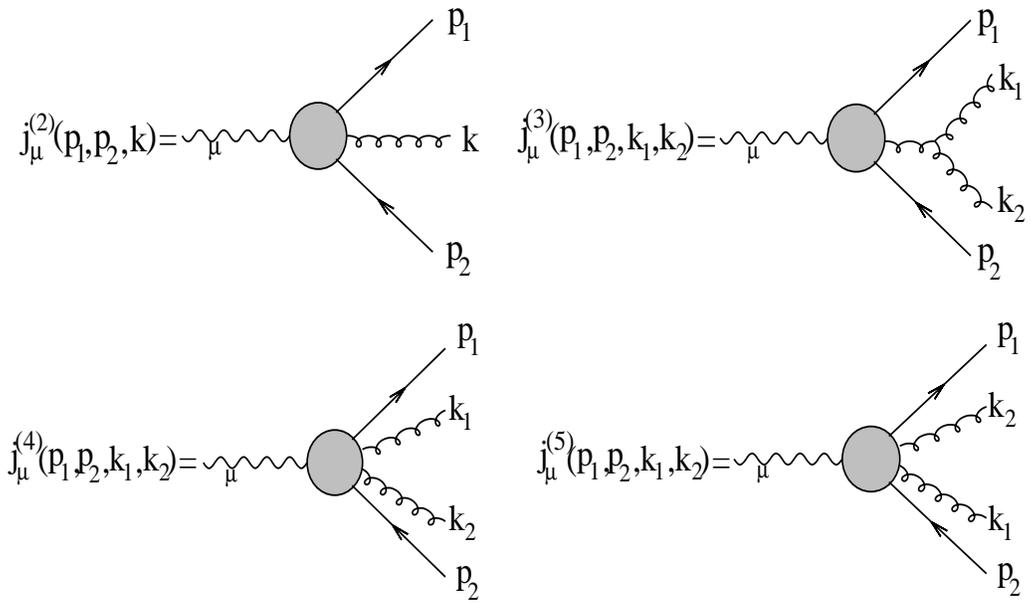,height=8cm,width=13.5cm}
\caption{Additional components needed to calculate the doubly resonant 
contributions to $e^{+}e^{-}\rightarrow q \bar{q} q \bar{q} gg$.  The shaded
blob represents a sum over possible 
attachments of gluons to the spinor line which preserve the order of
attachments.}
\end{figure}

\begin{equation}
\mathcal{M}(r_{1},r_{2};p_{1},p_{2},p_{3},p_{4})=
T^{\mu \nu}(r_{1},r_{2};p_{1}+p_{2},p_{3}+p_{4}) 
j^{(1)}_{\mu}(p_{1},p_{2}) j^{(1)}_{\nu}(p_{3},p_{4})
\end{equation}
where the terms are defined in fig 4 and the momentum labels refer to fig 1.
The computational complexity is reduced by assuming
massless electrons and massless quarks which has 
been done throughout this paper.

In this way all of the amplitudes can be built up from just the production
tensor $T^{\mu \nu}$ and some `decay currents' $j^{(i)}_{\mu}$.  In order to 
calculate all amplitudes for double gluon radiation four additional decay
currents (fig 5) are needed.  Note one must distinguish between
$j^{(4)}_{\mu}$ and $j^{(5)}_{\mu}$ which are related by $j^{(4)}_{\mu}=
j^{(5)}_{\mu}(k_{1} \leftrightarrow k_{2})$ in order to obtain the correct
colour factors.  

The decay currents may be contracted onto the production tensor to obtain 
eight distinct Lorentz-colour structures (colour-matrices are omitted
for clarity)

\begin{eqnarray}
\mathcal{M}_{1} & = & 
T^{\mu \nu}j^{(2)}_{\mu}(p_{1},p_{2},k_{1})j^{(2)}_{\nu}(
p_{3},p_{4},k_{2}) \nonumber \\
\mathcal{M}_{2} & = & \mathcal{M}_{1}
(k_{1}\leftrightarrow k_{2}) \nonumber \\
\mathcal{M}_{3a} & = & T^{\mu \nu}j^{(4)}_{\mu}(p_{1},p_{2},k_{1},k_{2})
j^{(1)}_{\nu}(p_{3},p_{4})  \nonumber \\
\mathcal{M}_{3b} & = & T^{\mu \nu}j^{(5)}_{\mu}(p_{1},p_{2},k_{1},k_{2})
j^{(1)}_{\nu}(p_{3},p_{4})  \nonumber \\
\mathcal{M}_{4a} & = & T^{\mu \nu}j^{(1)}_{\mu}(p_{1},p_{2})
j^{(4)}_{\nu}(p_{3},p_{4},k_{1},k_{2}) \nonumber \\
\mathcal{M}_{4b} & = & T^{\mu \nu}j^{(1)}_{\mu}(p_{1},p_{2})
j^{(5)}_{\nu}(p_{3},p_{4},k_{1},k_{2}) \nonumber \\
\mathcal{M}_{5} & = & T^{\mu \nu}j^{(3)}_{\mu}(p_{1},p_{2},k_{1},k_{2})
j^{(1)}_{\nu}(p_{3},p_{4}) \nonumber \\
\mathcal{M}_{6} & = & T^{\mu \nu}j^{(1)}_{\mu}(p_{1},p_{2})
j^{(3)}_{\nu}(p_{3},p_{4},k_{1},k_{2})
\end{eqnarray}
then the total production amplitude is (suppressing colour matrices)

\begin{equation} 
\mathcal{M}=\mathcal{M}_{1}+\mathcal{M}_{2}+\mathcal{M}_{3a}+
\mathcal{M}_{3b}+\mathcal{M}_{4a}+\mathcal{M}_{4b}+
\mathcal{M}_{5}+\mathcal{M}_{6}
\end{equation}
after squaring and summing over colours it is convenient to separate the
squared matrix element into different parts depending on the form of the
Breit-Wigner resonances.  In this way one finds six distinct terms

\begin{equation}
\sum_{colours}|\mathcal{M}|^{2}=M_{1}+M_{2}+M_{3}+M_{4}+M_{5}+M_{6}
\end{equation}
where
\begin{eqnarray}
M_{1} & = & N_{C}^{2}C_{F}^{2}\mathcal{M}_{1}\mathcal{M}_{1}^{*}\\
M_{2} & = & M_{1}(k_{1}\leftrightarrow k_{2})\\
M_{3} & = & N_{C}^{2}C_{F}^{2}(\mathcal{M}_{3a}\mathcal{M}_{3a}^{*}
+\mathcal{M}_{3b}\mathcal{M}_{3b}^{*})
+N_{C}^{3}C_{F}\mathcal{M}_{5}\mathcal{M}_{5}^{*} - \nonumber \\
& & \frac{1}{2}N_{C}C_{F}(\mathcal{M}_{3a}\mathcal{M}_{3b}^{*}+ 
\mathcal{M}_{3b}\mathcal{M}_{3a}^{*})+
 \nonumber \\ & &
\frac{1}{2}N_{C}^{3}C_{F}
([\mathcal{M}_{3a}\mathcal{M}_{5}^{*}+\mathcal{M}_{5}
\mathcal{M}_{3a}^{*}]-[\mathcal{M}_{3b}\mathcal{M}_{5}^{*}+
\mathcal{M}_{5}\mathcal{M}_{3b}^{*}])
\\
M_{4} & = & N_{C}^{2}C_{F}^{2}(\mathcal{M}_{4a}\mathcal{M}_{4a}^{*}+
\mathcal{M}_{4b}\mathcal{M}_{4b}^{*})
+N_{C}^{3}C_{F}\mathcal{M}_{5}\mathcal{M}_{5}^{*} - \nonumber \\  & &
\frac{1}{2}N_{C}C_{F}(\mathcal{M}_{4a}\mathcal{M}_{4b}^{*}+
\mathcal{M}_{4b}\mathcal{M}_{4a}^{*})+
 \nonumber \\ & &
\frac{1}{2}N_{C}^{3}C_{F}([\mathcal{M}_{4a}\mathcal{M}_{5}^{*}+
\mathcal{M}_{5}\mathcal{M}_{4a}^{*}]-[\mathcal{M}_{4b}\mathcal{M}_{5}^{*}+
\mathcal{M}_{5}\mathcal{M}_{4b}^{*}])
\\
M_{5} & = & \frac{1}{2}N_{C}C_{F}(\mathcal{M}_{1}\mathcal{M}_{2}^{*}+
\mathcal{M}_{2}\mathcal{M}_{1}^{*})\\
M_{6} & = & \frac{1}{2}N_{C}C_{F}[(\mathcal{M}_{3a}+\mathcal{M}_{3b})
(\mathcal{M}_{4a}+\mathcal{M}_{4b})^{*}+
\nonumber \\ & &
(\mathcal{M}_{4a}+\mathcal{M}_{4b})(\mathcal{M}_{3a}+\mathcal{M}_{3b})^{*}]
\end{eqnarray}

In the above expression the terms $M_{1},M_{2},M_{3},M_{4}$ denote the
unreconnected parts, while $M_{5}$ and $M_{6}$ correspond to interference
between the two decays.  The separation into reconnected and unreconnected
parts is (QCD) gauge invariant.

One can find small regions of phase space where the interference contribution
to the total transition probability is as large as 10\%, even for relatively 
energetic gluons ($\geq 5 \mbox{ GeV}$).  It is therefore 
not impossible that certain 
6 jet distributions could be significantly distorted by perturbative 
reconnection.

A Monte Carlo program was written to generate WW events with the
$q\bar{q}q\bar{q}gg$ final state using the Multichannel approach
\cite{16} with 24 channels based on the kinematic properties of
the contributing diagrams.  Events were generated at a centre-of-mass
energy of $192 \mbox{ GeV}$, although reconnection effects are expected
to be insensitive to the centre-of-mass energy within the LEP II range.
In addition specific phase space parameterisations
were computed which allowed efficient integration of both types of 
interference term.

Six jet final states were defined according to a minimum invariant mass between
partons.  A lower limit of $s_{cut}=1\mbox{ GeV}^{2}$ was used.  Strictly 
speaking this is too small for fixed order perturbation theory to be 
applicable, however the philosophy is that the results obtained will
provide an upper limit on reconnection effects, since 
moving to larger values
of $s_{cut}$ generally reduces any effect.  The six partons were clustered to
four jets using the Durham algorithm\cite{17}.  
Distributions for the Bengtsson-Zerwas
angle\cite{18} $(\chi_{BZ})$,
the modified Nachtmann-Reiter angle\cite{19} $(\theta_{NR})$ 
and the angle between the two lowest energy
jets $(\alpha_{34})$ 
were computed with and without the interference terms.
These angles are defined by equation 15 below
in which $\mathbf{p_{1},p_{2},p_{3}}$ and $\mathbf{p_{4}}$ are the
energy ordered jet 3-momenta.  

\begin{eqnarray}
\cos(\chi_{BZ}) & = & \mathbf{\frac{(p_{1}\times p_{2})\cdot
(p_{3} \times p_{4})}{|p_{1}\times p_{2}||p_{3} \times p_{4}|}}
\nonumber \\
\cos(\theta_{NR}) & = & \mathbf{\frac{(p_{1}-p_{2})\cdot (p_{3}-p_{4})}
{|p_{1}-p_{2}||p_{3}-p_{4}|}}
\end{eqnarray}

With four jets 
there are three ways of pairing them.  For each pairing the average of the
invariant masses was computed.  Thus for each event one has three mass 
values corresponding to each of the three possible pairings.  The mass 
closest to the 
input W mass was chosen as the mass estimate (this is only one
of several possibilities suggested in \cite{5}).
Distributions for the mass calculated in this way were also produced with
and without colour interference terms.

The difference between distributions with and without the interference terms
was computed.  This shows the distortion induced by the interferences.

Finally the integral of the absolute value of the interferences was found for
gluon energies greater than $2 \mbox{ GeV}$, $5 \mbox{ GeV}$ and 
$10 \mbox{ GeV}$.  These quantities are finite since the interference
terms contain no collinear singularities (apart from integrable ones
when three partons become collinear), and provide an indication of the
possible size of interference effects in events with jet energies greater
than $2,5$ and $10 \mbox{ GeV}$.

\section{Results}

The mean mass can be calculated with and without the reconnected terms.  The 
result one finds depends on the choice of invariant mass cut, but must tend
to zero as $s_{cut}\rightarrow 0$ since the unreconnected terms are more 
singular than the reconnected terms in this limit.  Mass shifts for a variety
of invariant mass cuts on the final state are shown in the table below.

\begin{center}
\begin{tabular}{|c|c|c|c|c|} \hline
 $s_{cut}/\mbox{GeV}^{2}$ & $0.1$ & $1.0$ & $10.0$ & $100.0$ \\\hline
 $\delta M_{W}/\mbox{MeV}$ & $-0.030$ &
$-0.045$ & $-0.025$ & $\sim -0.015$ \\\hline
\end{tabular}
%\caption{The estimated mass shift at lowest order perturbation theory for
%six jet events as defined by the minimum invariant mass $s_{cut}$.  The mass
%shifts are to the nearest 5 keV}
\end{center}
The exact numbers are also
slightly dependent on the reconstruction scheme used
for defining the experimental W mass.

\begin{figure}
\centering
\epsfig{bbllx=19,bblly=143,bburx=575,bbury=699,
file=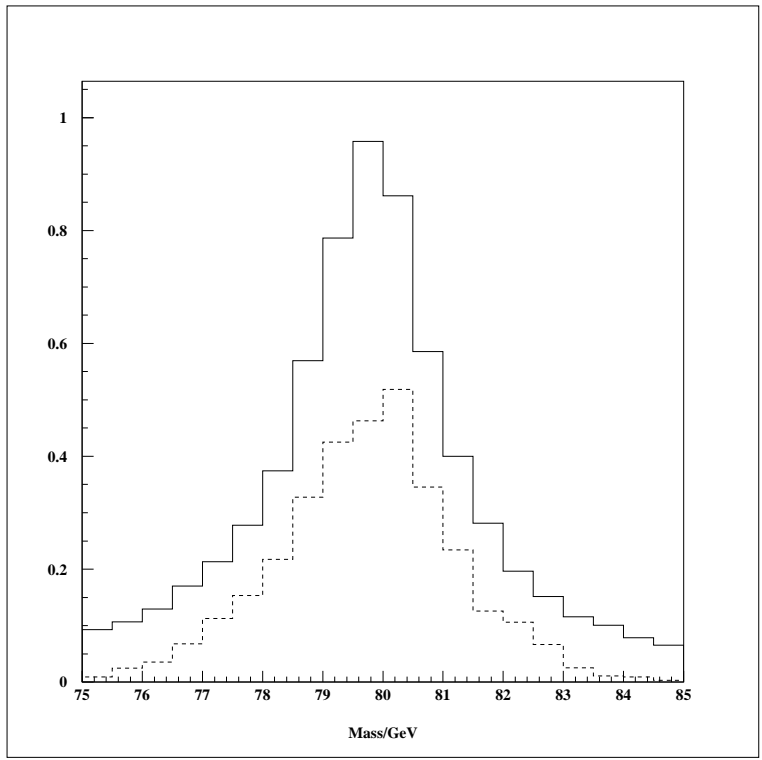,height=10cm}
\caption{Mass distribution for unreconnected events (solid line), change
induced by reconnected terms $\times 1000$ (dashed line), in arbitrary units}
\end{figure}

\begin{figure}
\centering
\epsfig{bbllx=19,bblly=143,bburx=575,bbury=699,
file=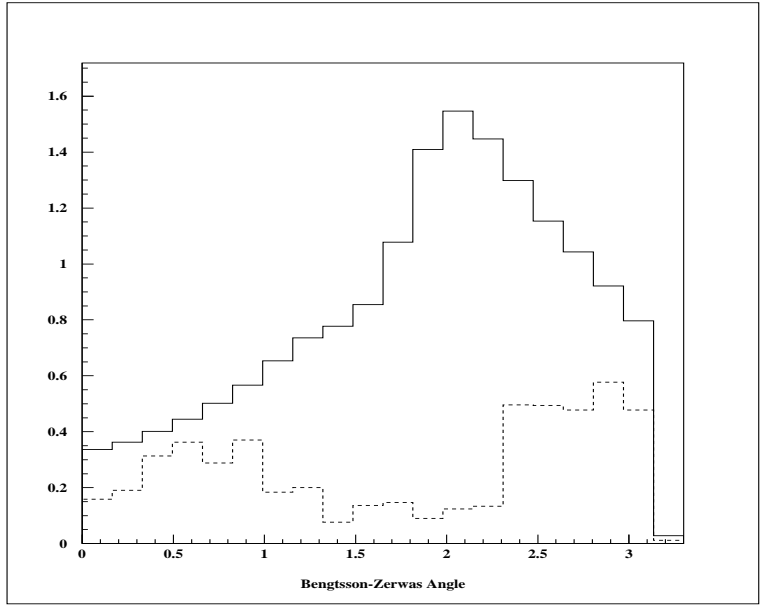,height=8cm,width=10cm}
\caption{Distribution of $\chi_{BZ}$, the Bengtsson-Zerwas Angle for 
unreconnected events (solid line), change induced by reconnected terms
$\times 1000$ (dashed line)}
\end{figure}

\begin{figure}
\centering
\epsfig{bbllx=19,bblly=143,bburx=575,bbury=699,
file=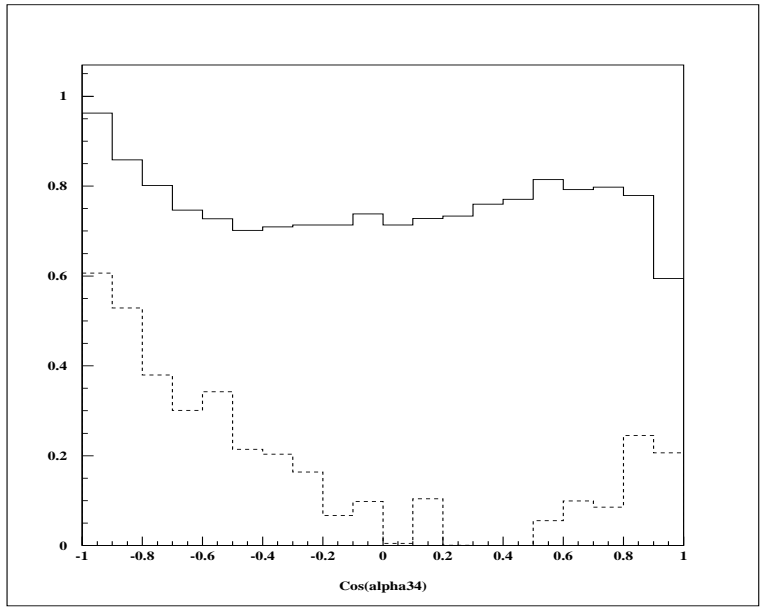,height=8cm,width=10cm}
\caption{Distribution of $\cos (\alpha_{34})$ for unreconnected events 
(solid line), change induced by reconnected terms $\times 500$ 
(dashed line), in arbitrary units}
\end{figure}

\begin{figure}
\centering
\epsfig{bbllx=19,bblly=143,bburx=575,bbury=699,
file=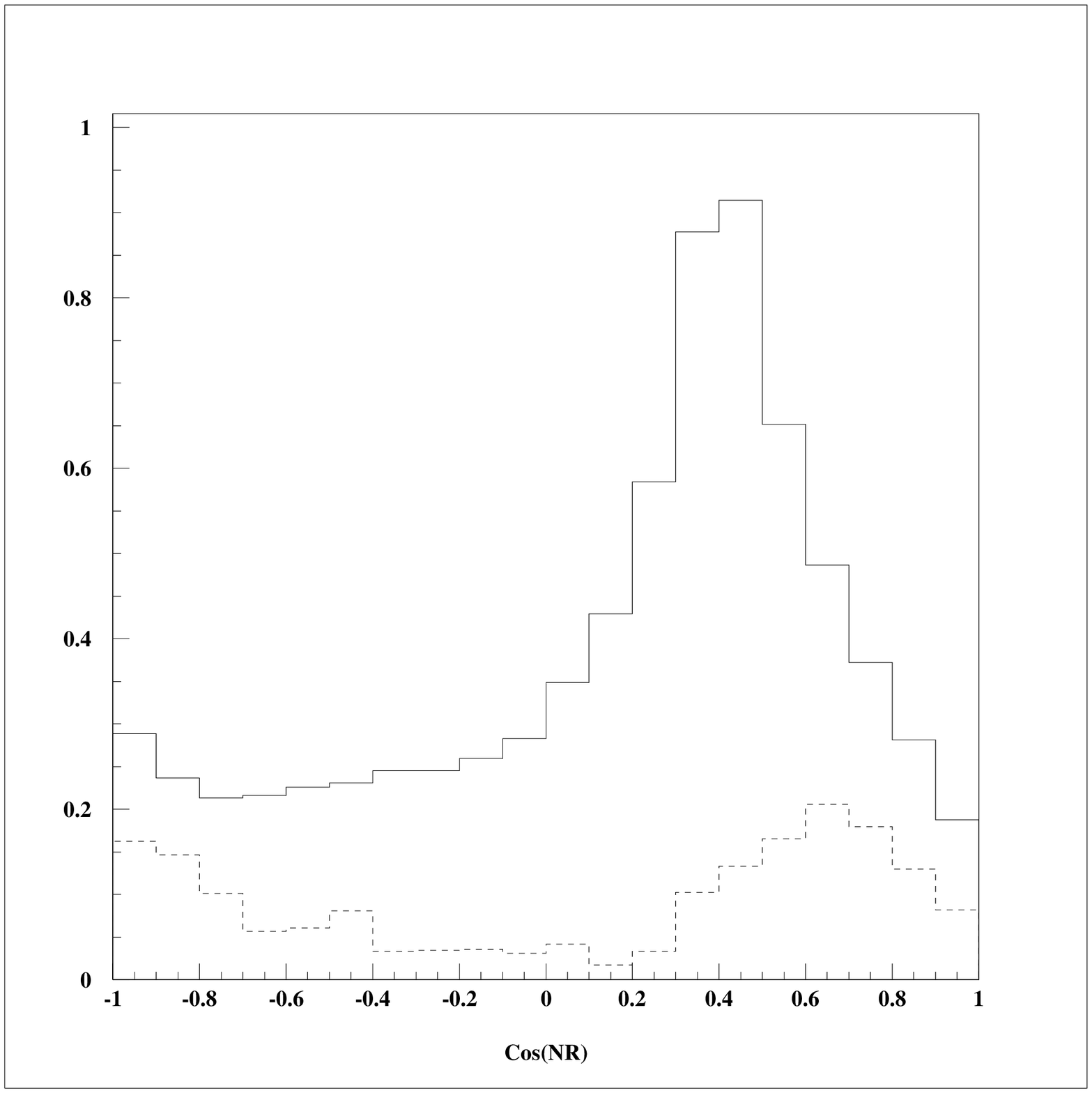,height=8cm,width=10cm}
\caption{Distribution of $\cos(\theta_{NR})$, the Nachtmann-Reiter
Angle for unreconnected events (solid line), change induced by 
reconnected terms $\times 500$ (dashed line)}
\end{figure}

Figure 6 shows the distribution of reconstructed mass using only 
the unreconnected parts of the matrix element (solid line).  The dashed line
shows one thousand times the change induced when the reconnected 
terms are present.  The distributions for the mass under the full matrix 
element and unreconnected terms only differ essentially by a multiplicative
constant of order $1.001$.  The mean value of the mass distribution is shifted
by less than a part per million due to the presence of reconnected terms.
 
Figures 7,8 and 9 show similar plots for the distribution of the
Bengtsson-Zerwas angle, the angle between the
two lowest energy jets $\alpha_{34}$ and the Nachtmann-Reiter angle.
It will be seen that the effect is at
or below the per mille level and is essentially just multiplicative, 
distortions
of the distributions occur at a much lower level.
These effects can be understood within the soft interference limit.  In the 
soft limit one may describe gluon radiation using eikonal vertices and the 
matrix element squared becomes.

\begin{equation}
|\mathcal{M}|^{2}=|\mathcal{M}_{0}|^{2}(H(k_{1},k_{2})+AG(k_{1})G(k_{2}))
\end{equation}
where $H(k_{1},k_{2})$ is the soft unreconnected distribution, 
A is some constant that will depend on the energy resolution and 
W width.
$G(k)$ is the reconnected distribution (note that at this order
the reconnected gluons are radiated independently, however this is not
true in higher orders) and is given by

\begin{equation}
G(k)=\frac{(p_{1} \cdot p_{4})}{(p_{1} \cdot k)(p_{4} \cdot k)}+
     \frac{(p_{2} \cdot p_{3})}{(p_{2} \cdot k)(p_{3} \cdot k)}-
     \frac{(p_{1} \cdot p_{3})}{(p_{1} \cdot k)(p_{3} \cdot k)}-
     \frac{(p_{2} \cdot p_{4})}{(p_{2} \cdot k)(p_{4} \cdot k)}
\end{equation}

One may integrate over the directions of each emission to find the 
enhancement due to soft interference between decays:

\begin{equation}
|\mathcal{M}_{rec}|^{2} \sim |\mathcal{M}_{0}|^{2}\times 
\ln^{2}\left(\frac{(p_{1} \cdot p_{4})(p_{2} \cdot p_{3})}
{(p_{1} \cdot p_{3})(p_{2} \cdot p_{4})}\right)
\end{equation}
where the momenta are as defined in fig 1.  

The effect of the reconnection terms is essentially to enhance coplanar
configurations where some invariant masses can be much larger than others.
In configurations where the W decay planes are at right angles, none of the 
parton directions can become close and so the argument of the logarithm in 
equation (18) is close to one and there is little enhancement. In most
approximately coplanar configurations the BZ angle will be close to either
$0$ or $\pi$ as both $\mathbf{p_{1}\times p_{2}}$ and $\mathbf{p_{3}
\times p_{4}}$ (energy ordered momenta) 
are likely to point out of the decay plane and hence be
either parallel or anti-parallel.  Thus one expects enhancement around these
values.

The situation for $\cos(\alpha_{34})$ is not quite so straight forward.
A similar argument favours $\cos(\alpha_{34}) \sim 1$ however this
configuration is suppressed by the jet reconstruction kinematics; one would
need two low energy quarks and both gluons radiated in approximately the
same direction \emph{and} to be clustered as two distinct jets.  However 
the configurations corresponding to $\cos(\alpha_{34}) \sim -1$  can be 
enhanced (see fig 10).

A similar argument for $\theta_{NR}$ is not so apparent as its geometrical 
interpretation is less clear (the angle between the axis defined by the
vector between the two lowest energy jets and that between the two highest
energy jets).  One may construct the enhancement due to equation (18) and 
find qualitatively the same shape as observed in figure 9.

\begin{figure}
\centering
\epsfig{
file=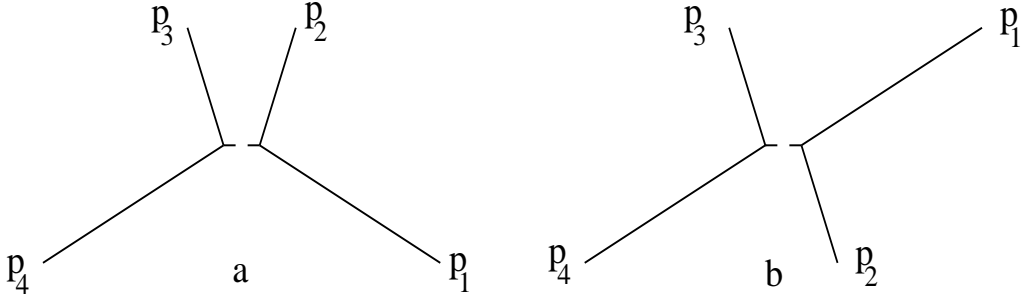,height=13.5cm,width=4cm,angle=270}
\caption{Kinematic configurations which are a) Enhanced by interference
but kinematically suppressed  
($\cos(\alpha_{34}) \sim 1$), and b) Enhanced by interference 
($\cos(\alpha_{34}) \sim -1$), since $(p_{1} \cdot p_{4})(p_{2} \cdot p_{3})
> (p_{1} \cdot p_{3})(p_{2} \cdot p_{4})$.
The momentum labels refer to fig 1,
and in this example  $p_{2}$ and $p_{3}$ correspond to the lowest energy
jets.}
\end{figure}

The absolute value of the interference terms was integrated over the region
defined by $\omega \geq 2 \mbox{ GeV}$, $\omega \geq 5 \mbox{ GeV}$ and
$\omega \geq 10 \mbox{ GeV}$ where $\omega$ is the minimum gluon energy.  This
was done for $s_{cut}=10,1.0,0.1,0.01 \mbox{ GeV}^{2}$ 
and illustrates collinear
finiteness (see results below).

\begin{center}
\begin{tabular}{|c|c|c|c|c|c|} \hline
\multicolumn{2}{|c|}{$ \sigma_{|int|} /\mbox{pb}$} & 
\multicolumn{4}{c|}{$s_{cut}/\mbox{GeV}^{2}$} \\ \cline{3-6}
\multicolumn{2}{|c|}{ } & $10$ & $1.0$ & $0.1$ & $0.01$ \\ \hline
 & 2 & 0.015 & 0.029 & 0.032 & 0.033 \\ \cline{2-6}
 $\omega /\mbox{GeV}$ & 5 & $0.0057$ & $0.0082$ &
$0.009$ & $0.01$ \\ \cline{2-6}
 & 10 & $0.0016$ & $0.0023$ & $0.0026$ &
$0.0026$ \\ \hline
\end{tabular}
\end{center}

The errors on these numbers are around 4\% each. Note that these are the 
integrated absolute value of the interference terms,  the actual contribution
of the interference terms to the cross-section is typically an order of
magnitude smaller due to large cancellations.

\section{Conclusions}

Effects of perturbative reconnection are not necessarily small, however the
regions of phase space in which sizable effects can occur \emph{are} small.  
Most experimentally interesting distributions are unaffected by reconnection 
at the perturbative level apart from a multiplicative factor close to unity.
In particular the mass distribution is shifted by 
less than one part per million 
by lowest order reconnection effects in 6-jet events.
Distributions sensitive to soft momenta seem to show 
greater distortion, however these effects are well 
below the per mille level 
and so unlikely to be seen at LEP II. 

The integration of the absolute value of the reconnection terms for gluon
energies above $5 \mbox{ GeV}$ shows that the maximum effect could only be
equivalent to a few events at LEP II and can probably be neglected
at this level of statistics.

Of course reconnection effects summed over higher terms, or within the 
hadronization phase need not be negligible and these effects still need to 
be addressed.

\section*{Acknowledgements}

Thanks must go to Dr. B. R. Webber for suggesting this topic and Dr. D. 
Summers for
many useful discussions.


\begin{thebibliography}{99}

\bibitem{1} Z. Kunszt, W.J. Stirling  et al., 
CERN-96-01 (1996) Vol1 p141 

\bibitem{2} W. Beenakker, A. P. Capovsky and F. Berends,  
Phys. Lett. \textbf{B411} (1997) 203

\bibitem{3} A. Denner, S. Dittmaier and M. Roth,  
Nucl. Phys. \textbf{B519} (1998) 39

\bibitem{4} L. Lonnblad and T. Sjostrand, 
Phys. Lett. \textbf{B351} (1995) 293

\bibitem{5} J. Hakkinen and M. Ringner,  
Eur. Phys. J. \textbf{C5} (1998) 275

\bibitem{6} V. Kartvelishvili, R. Kvatadze and R. Moller,
Phys. Lett. \textbf{B408} (1997) 331

\bibitem{7} T. Sjostrand and V. A. Khoze,  
Z. Phys. \textbf{C62} (1994) 281

\bibitem{8} T. Sjostrand and V. A. Khoze,
Phys. Rev. Lett. \textbf{72} (1994) 28

\bibitem{9} C. Friberg and G. Gustafson and J. Hakkinen,  
Nucl.Phys. \textbf{B490} (1997) 289

\bibitem{10} J. Ellis and K. Geiger,
Phys. Rev. \textbf{D 54} (1996) 1967

\bibitem{11} P. B. Renton, W. J. Stirling, D. R. Ward et al.,
J. Phys. G:Nucl. Part. Phys. \textbf{24} (1998) 365

\bibitem{12} E. Accomando, A. Ballestrero and E. Maina,
Phys. Lett. \textbf{B362} (1995) 141

\bibitem{13} V. S. Fadin, V. A. Khoze and A. D. Martin,  
Phys. Rev.  \textbf{D 49} (1994) 2247

\bibitem{14} V. S. Fadin, V. A. Khoze and A. D. Martin,  
Phys. Lett. \textbf{B320} (1994) 141 

\bibitem{15} A. Ballestrero and E. Maina, 
Phys. Lett. \textbf{B350} (1995) 225

\bibitem{16} R. Kleiss and R. Pittau,
Comput. Phys. Commun. \textbf{83} (1994) 141

\bibitem{17} S. Catani, Y. L. Dokshitser, M. Olsson, G. Turnock and
B. R. Webber, Phys. Lett. \textbf{B269} (1991) 432

\bibitem{18} M. Bengtsson and P. M. Zerwas,
Phys. Lett. \textbf{B208} (1998) 306

\bibitem{19} S. Bethke, A. Richter and P. M. Zerwas, Z. Phys.
\textbf{C49} (1991) 59


\end{thebibliography}
\end{document}